\documentclass{article}
\usepackage{spconf,amsmath,graphicx,hyperref}
\usepackage{xcolor}
\usepackage{amssymb}
\usepackage{bbding}
\usepackage{booktabs}
\usepackage{pifont}
\usepackage{makecell}
\usepackage{graphicx}
\usepackage{subcaption}
\usepackage[margin=1in]{geometry}
\usepackage{graphicx}
\usepackage{subcaption}
\usepackage{multirow}
\usepackage[font=small,labelfont=bf]{caption}
\newcommand{\cmark}{\ding{51}} 
\newcommand{\xmark}{\ding{55}} 

\def\Ndim{{\textit{$\mathcal{N}$-dimension}}}
\def\D{{$\mathcal{D}$ }}
\def\F{{$\mathcal{F}$ }}
\def\N{{$\mathcal{N}$ }}

\newcommand{\metric}{\mathrm{Metric}_{\mathrm{IF}}}

\title{ISA-Bench: Benchmarking Instruction Sensitivity for Large Audio Language Models}
\name{\shortstack[c]{Bohan Li$^{*}$, Wenbin Huang$^{*}$, Yuhang Qiu$^{*}$, Yiwei Guo, Hankun Wang, Zhihan Li, \\ Jing Peng, Ziyang Ma, Xie Chen, Kai Yu$^{\dagger}$ \thanks{ \scriptsize{$^*$ means equal contribution,} $^\dagger$\scriptsize{is the corresponding author.}}}}
\address{
X-LANCE Lab, School of Computer Science,  Shanghai Jiao Tong University, China \\MoE Key Lab of Artificial Intelligence; Jiangsu Key Lab of Language Computing, China \\
\texttt{\{everlastingnight, kai.yu\}@sjtu.edu.cn}
}
\begin{document}
\ninept
\maketitle
\begin{abstract}
Large Audio Language Models (LALMs), which couple acoustic perception with large language models (LLMs) to extract and understand diverse information from audio, have attracted intense interest from both academic and industrial communities. 
However, existing LALMs are highly sensitive to how instructions are phrased, affecting both (i) instruction-following rates and (ii) task performance.  Yet, no existing benchmarks offer a systematic and comprehensive evaluation of this sensitivity. We introduce \textbf{ISA-Bench}, a dynamic benchmark evaluating instruction sensitivity for LALMs along three axes: instruction description, output format, and task composition. We assess recent open-source and proprietary LALMs using ISA-Bench, profiling both compliance and accuracy under controlled instruction variations. Experimental results reveal that even state-of-the-art LALMs suffer significant instruction sensitivity, leading to degraded performance on fundamental audio understanding tasks. To mitigate this issue, we fine-tune Qwen2-Audio on a specifically constructed complex instruction‑variant dataset, achieving a marked improvement in instruction-following performance. However, this also induces nontrivial catastrophic forgetting: the model loses some previously mastered task capabilities when exposed to new instruction styles. Our benchmark provides a standardized basis for assessing and improving instruction sensitivity in LALMs, underscoring the need for instruction-robust audio understanding in real-world pipelines. \footnote{\scriptsize{https://github.com/bovod-sjtu/ISA-Bench}}

\end{abstract}
\section{Introduction}
Audio is a core modality for human-computer interaction. Recent advances have empowered large language models (LLMs) with audio perception ability by adding neural encoding layers, producing large audio-language models (LALMs) that can handle universal audio understanding tasks given audio signals and textual instructions~\cite{peng2024survey}. 
In this paradigm, instructions are essential: they define what should be extracted from the audio, the reasoning to be applied, and the form of output required.

In NLP, prior work has shown that the format and phrasing of instructions or prompts strongly affect LLM performance \cite{sclar2024quantifying}. Benchmarks and optimization methods have been developed to assess and improve this instruction-following ability~\cite{zhou2023instruction, qin2024infobench, wen2024bench, dong2025selfplay,an2025ultraifadvancinginstructionfollowing}. However, LALMs face an extra challenge: beyond understanding the instruction text, they must also perceive information from audio, making it harder to satisfy both instruction compliance and task accuracy. Moreover, published evaluations of LALMs mostly use instruction forms that are seen during supervised fine-tuning (SFT), giving an upper‐bound of performance estimate. In real deployment, models will encounter unseen instruction variants, and performance will typically degrade under such scenarios.

Consequently, the notion of instruction sensitivity has recently been introduced and recognized as a critical challenge for LALMs~\cite{peng2024survey, lu2025speech,guo2025ahamask}: these models are expected not only to follow the instructions but also to maintain strong task performance. As summarized in Table ~\ref{tab:bench_comparison}, existing benchmarks typically address only one aspect of instruction sensitivity, whereas such aspects should in fact be considered holistically to evaluate the capacity of LALMs as intelligent agents in universal audio understanding tasks. To this end, we propose \textbf{ISA-Bench} (\textbf{I}nstruction \textbf{S}ensitivity of large \textbf{A}udio language models \textbf{Bench}mark), a multidimensional and dynamic benchmark designed to comprehensively assess the instruction sensitivity of LALMs.
More specifically, the proposed benchmark is organized along three principal dimensions: (1) the \textit{$\mathcal{D}$-dimension}, which concerns the textual description and phrasing of instructions; (2) the \textit{$\mathcal{F}$-dimension}, which evaluates compliance with output format requirements; and (3) the \textit{$\mathcal{N}$-dimension}, which measures the number of subtasks composed within a single instruction. We evaluate several state-of-the-art LALMs, including both open-source and proprietary systems, across five atomic tasks: automatic speech recognition (ASR), speech-to-text translation (S2TT, English-to-Mandarin), speech emotion recognition (SER), gender recognition (GR), and audio captioning (AAC). For each dimension and task, we dynamically set the best-achieved score among all models as the reference performance, and then assign each model a relative score with respect to this reference. To ensure diversity in the evaluation set, we generate instruction variants via LLM-based rewriting and recomposition across multiple phrasings and formatting styles. Likewise, we assemble a multi-task instruction corpus for SFT under the same diversity. 

Our experiments demonstrate: (i) instruction sensitivity remains an unresolved challenge for LALMs: even state-of-the-art models still degrade significantly under varied instruction forms; (ii) SFT remains insufficient: fine-tuning with diverse instructions can improve instruction following ability, but it often leads to catastrophic forgetting on mastered tasks. This finding highlights the inherent difficulty of the instruction sensitivity problem and suggests that more sophisticated solutions are required. 
We anticipate that ISA-Bench will encourage researchers to explore improved approaches to enable LALMs to deliver more robust and reliable outputs under real-world instruction scenarios.

\vspace{-0.5em}
\begin{table}[t] 
\centering
\caption{Comparison of instruction-related benchmarks. }
\label{tab:bench_comparison}
\setlength{\tabcolsep}{4pt} 
\renewcommand{\arraystretch}{1.2} 
\small 
\resizebox{\linewidth}{!}{
\begin{tabular}{l|ccc}
\toprule
\textbf{Benchmark} & \makecell{\textbf{Instruction} \\ \textbf{Following}} 
                   & \makecell{\textbf{Performance} \\ \textbf{Robustness}} 
                   & \makecell{\textbf{Composite} \\ \textbf{Tasks}} \\
\midrule
IFEval-Audio~\cite{gao2025ifevalaudio}  & \cmark & \xmark & \xmark \\
Speech-IFEval~\cite{lu2025speech} & \cmark & \xmark & \xmark \\
\textbf{ISA-Bench \textit{(ours)}}            & \cmark & \cmark & \cmark \\
\bottomrule
\end{tabular}}
\vspace{-10pt}
\end{table}

\section{Related Work}

\textit{\textbf{Large Audio Language Models}}\quad LALMs commonly pair a pretrained LLM backbone with an audio front end. Typically, an audio encoder first produces acoustic representations; a lightweight projector layer then maps these features into the LLM’s embedding space. Training is typically end-to-end, fine-tuning the encoder, projector, and often the LLM backbone. While implementations vary in module choices and optimization strategies, the overall architecture is mainly consistent. These models have shown strong scalability and performance on universal audio understanding tasks~\cite{tang2024salmonn, chu2024qwen2audio, xu2025qwen25omnitechnicalreport, hu-etal-2024-wavllm, lu24c_interspeech_desta, lu2025desta25audiogeneralpurposelargeaudio, ding2025kimi, microsoft2025phi4minitechnicalreportcompact}.

\noindent\textit{\textbf{Instruction Sensitivity.}}\quad We use instruction sensitivity to denote how a model’s output depends on both (i) instruction following ability and (ii) task performance robustness.
In NLP, instruction following has been extensively studied using dedicated benchmarks and training strategies~\cite{zhou2023instruction, qin2024infobench, wen2024bench, dong2025selfplay,an2025ultraifadvancinginstructionfollowing}. On the other hand, the prompt sensitivity is observed: with fixed task requirements, small changes in instruction wording can alter model behavior and reduce task performance~\cite{sclar2024quantifying, worstprompt_nips2024, zhuo-etal-2024-prosa}. In the audio domain, benchmarks such as Speech-IFEval~\cite{lu2025speech} and IFEval-Audio~\cite{gao2025ifevalaudio} primarily assess compliance. However, as summarized in Table~\ref{tab:bench_comparison}, existing benchmarks largely omit a key dimension of instruction sensitivity: robustness of performance to instruction variation, which is crucial in practice. In addition, when LALMs are instructed with composite tasks, we observe pronounced sensitivity in response quality. The absence of these two factors motivate us to develop ISA-Bench.

\vspace{-5pt}
\section{ISA-Bench}
\vspace{-5pt}
\subsection{Benchmark Formulation}
\vspace{-5pt}

In typical scenarios, task-specific instructions are provided together with explicit requirements for the desired output format. Formally, the structure of such an instruction $\mathcal{I}$ can be expressed as:

\begin{equation}
\mathcal{I} = \mathcal{D}(\{t_i\}_{i=1}^\mathcal{N}, \mathcal{F}),
\end{equation}
where $\mathcal{D}$ denotes the textual description, $\mathcal{F}$ specifies the output format requirement, $\{\textbf{t}\}$ represents the set of subtasks to be executed by the LLM, and $\mathcal{N}$ is the total number of subtasks. 

Although LALMs incorporate acoustic information into the modeling process, their received instructions and expected outputs can be described in the same formal manner. At this stage, we explicitly decompose the universal instruction into three primary components, denoted by the symbols: $\mathcal{D}$, $\mathcal{F}$ and $\mathcal{N}$. Building upon these three components, we establish them as the core dimensions of the ISA-Bench framework, along which both the dataset construction and subsequent performance evaluation are carried out.

\begin{table}[t] 
\centering
\caption{Overview of tasks, evaluation metrics, test sets, and number of audio samples in ISA-Bench. ``*’’ refers to subset in IEMOCAP Session 5 having transcripts, emotion, and gender annotations.}
\vspace{-10pt}
\setlength{\tabcolsep}{0pt}
\renewcommand{\arraystretch}{1.15}
\small
\label{tab:datasets}
\resizebox{\linewidth}{!}{
\begin{tabular}{lccc c}
\toprule
\textbf{Tasks} & \textbf{Metrics} & \textbf{Test Sets} & \textbf{Num.\ Samples} \\ 
\midrule
\multicolumn{4}{l}{\textit{\textbf{Atomic Tasks}}} \\
ASR(Automatic speech recognition) & IFR, WER$_\text{IF}$ & LibriSpeech test-clean~\cite{librispeech} & 2620 \\
S2TT(en→zh, speech-to-text translation) & IFR, BLEU$_\text{IF}$~\cite{bleu} & CoVoST2 (en→zh) test~\cite{covost2} & 15531 \\
SER(speech emotion recognition) & IFR, ACC$_\text{IF}$ & IEMOCAP Session 5~\cite{iemocap} & 1241 \\
GR(gender recognition) & IFR, ACC$_\text{IF}$ & LibriSpeech test-clean & 2620 \\
AAC(automatic audio captioning) & IFR, METEOR$_\text{IF}$~\cite{meteor} & AudioCaps test~\cite{audiocaps} & 964 \\
\midrule
\multicolumn{4}{l}{\textit{\textbf{Composite Tasks}}} \\
\makecell[l]{2- or 3-way composition \\ of ASR, SER, and GR} & \makecell{IFR, WER$_\text{IF}$, \\ ACC$_\text{IF}$} & \makecell{IEMOCAP Session 5 subset$^*$} & 791 \\
\bottomrule
\end{tabular}}
\vspace{-10pt}
\end{table}
\vspace{-10pt}

\subsection{Tasks, task-native Metrics and Datasets}

We consider five atomic tasks: ASR, S2TT (English to Madarin, en$\rightarrow$zh), SER, GR and AAC, together probing the audio understanding capabilities of LALMs. We adopt task-native metrics--- WER for ASR, BLEU~\cite{bleu} for S2TT, accuracy (ACC) for SER and GR, and METEOR~\cite{meteor} for AAC--- and convert them to their compliance-aware counterparts, as described in Section~\ref{sec:eval_metrics}. As summarized in Table~\ref{tab:datasets}, five public datasets are employed to support the evaluation of different tasks. For $\mathcal{N}$-dimension, we construct composite tasks based on three atomic tasks: ASR, SER, and GR. We consider all perturbation of these 3 tasks to construct . These composite tasks require LALMs to perform two or three subtasks and generate their outputs sequentially in a specified format. A subset of IEMOCAP session 5 is adopted for evaluating composite tasks, as it simultaneously provides sufficient audio transcriptions($>$ 5 words), emotion labels, and gender annotations.

\vspace{-5pt}

\subsection{Construction of Instruction Variants}

To rigorously evaluate the instruction sensitivity of LALMs, it is essential to ensure sufficient diversity in the instructions employed. Across the three dimensions, the construction of instructions follows distinct design principles, as the evaluation objectives differ. The construction methods for each dimension are detailed as followed:

\noindent\textit{\textbf{$\mathcal{D}$-dimension}}\quad In this dimension, we focus on variations in the textual description of instructions. Following prior work \cite{peng2024survey}, we construct a diverse set of instruction variants that encompass alterations in punctuation, semantic complexity and case sensitivity. Beyond these, we further introduce two robustness-oriented variants, incorporating syntax errors and lexical errors respectively. We decompose an instruction into four fragments, formulated their concatenation as:

\begin{equation}
\label{eq:instruction}
    \mathcal{I}=\mathcal{D}_t(\{t_i\}_{i=1}^\mathcal{N})\mathcal{P}_c\mathcal{D}_f(\mathcal{F})\mathcal{P}_e
\end{equation}
where $\mathcal{D}_t$ denotes the task description, $\mathcal{D}_f$ specifies the output format description, $\mathcal{P}_c$, $\mathcal{P}_e$ represent the connecting and ending punctuations, respectively. For ASR, S2TT and AAC, we fix the output format specification $\mathcal{F}$ to require that model responses begin with the prefix ``The \{transcription\} / \{translation\} / \{audio caption\} is:". In contrast, for speech emotion recognition (SER) and gender recognition (GR), we constrain the response to a single word without any other content. Specifically, SER outputs are limited to one of \(\{\text{Happy}, \text{Sad}, \text{Angry}, \text{Neutral}\}\), while GR outputs are restricted to \(\{\text{Male}, \text{Female}\}\). A default instruction for an audio sample is represented by four base components: $\mathcal{D}_t$, $\mathcal{D}_f$, $\mathcal{P}_c$, and $\mathcal{P}_e$. We apply GPT-4 to rewrite specific components of the default instruction, yielding (i) case, semantic–complexity and robustness-oriented variants via $\{\mathcal{D}_t, \mathcal{D}_f\}$ and (ii) punctuation–style variants via $\{\mathcal{P}_c, \mathcal{P}_e\}$.

\noindent\textit{\textbf{$\mathcal{F}$-dimension}}\quad In this dimension, we focus on the output format requirements of instructions. We consider a range of common formats, including answer-only constraints, case sensitivity (upper and lower case, except for the S2TT task), prefix and suffix prompts, tag-wrapped outputs, and json-style formatting. According to Equation ~\ref{eq:instruction}, we only adjust $\mathcal{D}_f$ and $\mathcal{F}$ from the base instruction to construct variants. 

\noindent\textit{\textbf{$\mathcal{N}$-dimension}}\quad In this dimension, we focus on the number of subtasks contained within an instruction, aiming to assess LALMs' performance on composite tasks. We select three atomic tasks---ASR, SER, and GR---as candidate subtasks, and set the subtask number to either two or three. Since the ordering of subtasks may influence response quality, we evaluate instructions under all possible permutations of subtasks. Moreover, we adopt two distinct output formats: symbol-separated and JSON-style. All variants are instantiated from a unified template:

\begin{equation}
    \mathcal{I} = \{\mathcal{D}_{t_i}(t_i)\}_i^{\mathcal{N}}\mathcal{D}_{f}(\mathcal{F})
\end{equation}
where $\mathcal{D}_{t_i}(t_i)$ denotes the description of the $i$-th subtask, and $\mathcal{D}_{f}(\mathcal{F})$ refers to the formatting requirement. For a given audio sample, the descriptive styles $\mathcal{D}_{t_i}(t_i)$ and $\mathcal{D}_{f}$ remain fixed, while $\mathcal{N}$, $\mathcal{F}$ and the task order vary across different variants.

\vspace{-5pt}
\subsection{Evaluation Strategies and Compliance-aware Metrics}
\label{sec:eval_metrics}
\noindent\textit{\textbf{Instruction Following Evaluation}}
\quad We evaluate instruction sensitivity by first computing the instruction-following rate of outputs that satisfy the output-format constraints specified in the instructions. For most formatting requirements, compliance is verified with lightweight regular expressions. In the ASR setting, beyond format checks we enforce certain WER (100$\%$) and insertion error number ($\ge$3) as thresholds to flag off-spec responses (e.g., chit-chat, QA or unexpected prefixes) that violate the ASR output specification. For JSON-style outputs, we parse the response $\mathcal{R}$ with \texttt{json.loads($\mathcal{R}$)} (Python command) and treat successful parsing as a necessary condition for compliance. For answer-only constraints in open-ended tasks (ASR, S2TT, AAC), we use a small set of regex patterns and special-case rules, providing a competitive yet practical alternative to LLM-as-a-judge verification.

\noindent\textit{\textbf{Performance Robustness Evaluation and Scoring}}
\quad Similar to prior work~\cite{peng2024survey}, we report a compliance-aware task metric, $\mathrm{Metric}_{\mathrm{IF}}$, which credits task performance only when the response satisfies the required format: 
\vspace{-10pt}

\begin{equation}
\resizebox{\linewidth}{!}{$
\mathrm{Metric}_{\mathrm{IF}}(\mathcal{S})
=\frac{1}{|\mathcal{S}|}\sum_{i}\Bigl[
\mathrm{Metric}(g_i,h_i)\,\mathbf{1}_{\{h_i\in\mathcal{F}\}}
+\mathrm{Metric}(g_i,\varnothing)\,\mathbf{1}_{\{h_i\notin\mathcal{F}\}}
\Bigr]
$}
\end{equation}

where $\mathcal{S}=\{(g_i,h_i)\}$ denotes reference–hypothesis pairs, namely, $\mathcal{F}$ is the set of format-compliant outputs, and $\mathbf{1}_{\{\cdot\}}$ is the indicator function. $\mathrm{Metric}(\cdot,\cdot)$ is the base per-instance task metric. Thus, noncompliant hypotheses are evaluated as empty outputs, i.e., $\mathrm{Metric}(g_i,\varnothing)$.

For each task, we evaluate LALMs using $\mathrm{Metric}_{\mathrm{IF}}$ and report a Relative Performance to State-of-the-Art score(RPS). Following ~\cite{peng2024survey}, for higher-is-better metrics(e.g. BLEU~\cite{bleu}, ACC), we define: $RPS=\frac{\text{Model $\metric$}}{\text{SOTA $\metric$}}$. In contrast, for lower-is-better metrics(e.g. WER), we define: $RPS=\frac{\text{SOTA $\metric$}}{\text{Model $\metric$}}$.

\begin{figure*}[t] 
    \centering
    \includegraphics[width=\textwidth]{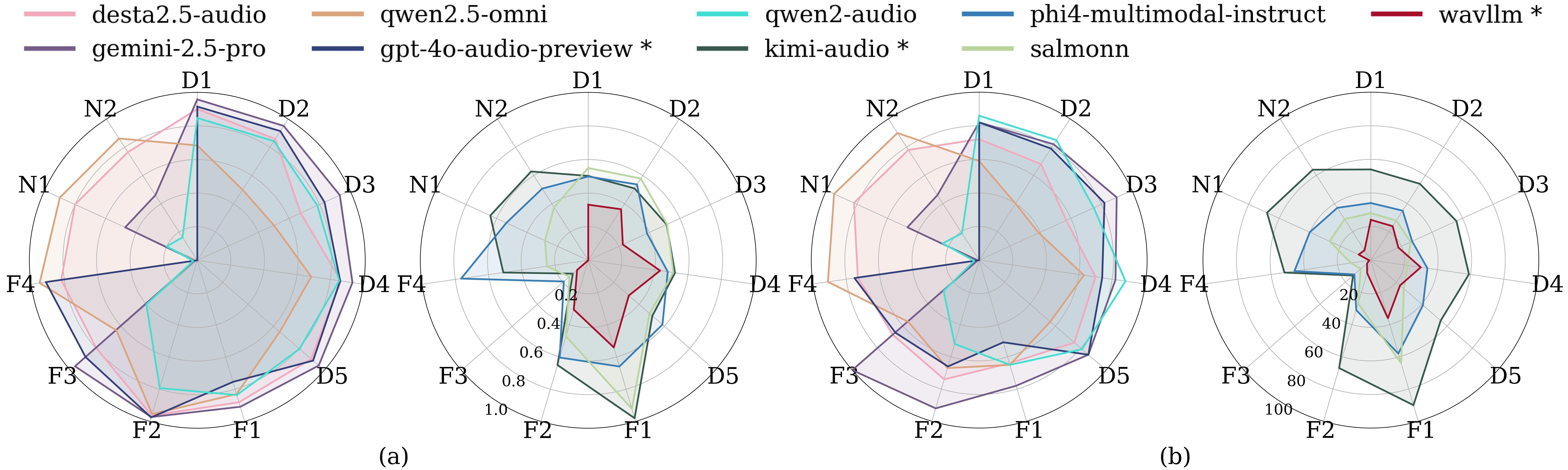} 
    \vspace{-20pt}
    \caption{Two radar plots in (a) show the average $\mathrm{IFR}$, and (b) presents the average RPS score across tasks. IDs refer to Table~\ref{tab:dim_variations}.}
    \label{fig:radar}
    \vspace{-20pt}
\end{figure*}

\vspace{-5pt}

\begin{figure}[t] 
    \centering
    \includegraphics[width=\linewidth]{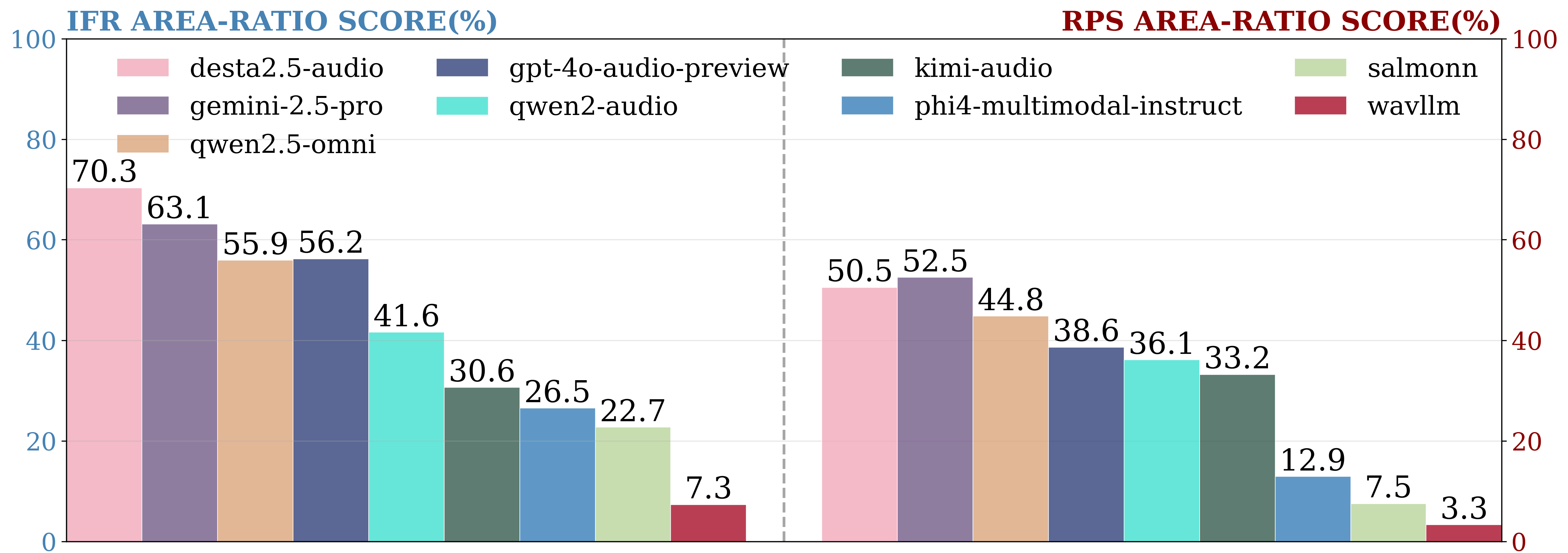} 
    \vspace{-16pt}
    \caption{Normalized radar plot areas of different models (maximum polygon area = 1). Left: IFR area; Right: RPS area.}
    \label{fig:area}
    \vspace{-20pt}
\end{figure}

\begin{table}[t]
  \centering
  \setlength{\tabcolsep}{5pt}
  \renewcommand{\arraystretch}{1.0}
  \scriptsize
  \begin{tabular}{@{}c c c c@{}}
    \toprule
    \textbf{Dimension} & \textbf{ID} & \textbf{Variation Class} & \textbf{Instruction variants} \\
    \midrule
    \multirow{5}{*}{\textbf{\D}} 
      & D1 & default             & default instruction \\
      & D2 & case                & upper case, lower case \\
      & D3 & robustness          & syntax error, lexical error \\
      & D4 & semantic complexity & simple, neutral, complex \\
      & D5 & punctuation         & punctuation alteration \\
    \midrule
    \multirow{4}{*}{\textbf{\F}} 
      & F1 & constrain           & answer-only constrain \\
      & F2 & case                & upper case, lower case \\
      & F3 & decoration          & prefix, suffix, tag-wrapped \\
      & F4 & json                & json-style formatting \\
    \midrule
    \multirow{2}{*}{\textbf{\N}} 
      & N1 & 2-task              & json, separator \\
      & N2 & 3-task              & json, separator \\
    \bottomrule
  \end{tabular}
  \caption{Mapping of dimension labels in Figure~\ref{fig:radar}, including variations and subclasses.}
  \label{tab:dim_variations}
  \vspace{-20pt}
\end{table}

\section{Experiments}
\vspace{-5pt}
\subsection{Experimental Settings}

\textit{\textbf{Tested Models}}\quad We benchmark nine recent LALMs, comprising two proprietary systems (GPT-4o-Audio~\cite{openai2024gpt4ocard}, Gemini 2.5 Pro~\cite{comanici2025gemini25pushingfrontier}) and seven open-source models (SALMONN-13B~\cite{tang2024salmonn}, WavLLM~\cite{hu-etal-2024-wavllm}, Qwen2-Audio-Instruct-7B~\cite{chu2024qwen2audio}, Qwen2.5-Omni-7B~\cite{xu2025qwen25omnitechnicalreport}, Kimi-Audio~\cite{ding2025kimi}, Phi-4-multimodal-instruct~\cite{microsoft2025phi4minitechnicalreportcompact}, and DeSTA2.5-Audio~\cite{lu2025desta25audiogeneralpurposelargeaudio}). Experiments are conducted on NVIDIA A800 GPUs.

\noindent\textit{\textbf{Variation of Dimensions}}\quad We perform a fine-grained partitioning of the three dimensions by their construction methods, spelling out subclasses of broadly defined variation categories. As shown in Table \ref{tab:dim_variations}, we map the IDs in Figure \ref{fig:radar} to concrete variation types, each corresponding to one or more instruction variants. For each variation, we aggregate the performance over its instruction variants and report the average score.

\vspace{-5pt}
\subsection{Evaluation Results}

\noindent\textit{\textbf{Overview Results}}\quad Figure \ref{fig:radar} reports both IFR and $\metric$ RPS score across every tested LALM. Note that marked ``$*$'' models have limitations: Kimi-Audio has no S2TT exposure, WavLLM lacks AAC data, and GPT-4o-Audio does not identify speaker gender. We exclude Kimi-Audio from S2TT evaluations, WavLLM from AAC, and GPT-4o-Audio from both the GR task and all \Ndim variants. We define a model’s total score as the average of its IFR and RPS over all atomic tasks. The results reveal that instruction sensitivity poses significant challenges. Gemini-2.5-Pro handles most variations in the \D and \F dimensions, but fails under JSON-style formatting constraints, resulting in degraded performance on variation F4 and all \N variants\footnote{\scriptsize{Area score can be raised to 80.0 by specially fixing the json-style responses.}}. GPT-4o-Audio is competitive; however, beyond failing to recognize speaker gender, it often produces commentary-like content rather than cleanly providing the task answer(variation F3). DeSTA2.5-Audio performs well overall, benefiting from its training design focused on instruction compliance. 

\noindent\textit{\textbf{Area-Ratio Score Reflecting Variation Robustness}}\quad Figure \ref{fig:area} presents normalized area ratios of radar plots as a summary measure of variation robustness. DeSTA2.5-Audio attains the highest overall instruction-following capability, while Gemini-2.5-Pro, despite its weakness under JSON formatting constraints, remains competitive. However, compared to Gemini-2.5-Pro, DeSTA2.5-Audio exhibits lower robustness on certain hard variation classes, resulting in lower RPS in those settings. It is also worth noting that Qwen-2.5-Omni shows relatively low instruction sensitivity, despite its omni-model nature. Still, all tested models leave considerable room for improvement: even the top performers achieve only about half of the maximum area-ratio. This suggests that no single model currently leads across all dimensions. \textit{We recommend that future evaluations adopt area-ratio scoring to comprehensively represent instruction sensitivity robustness.}

\begin{figure}[t]
    \centering
    \includegraphics[width=0.95\columnwidth]{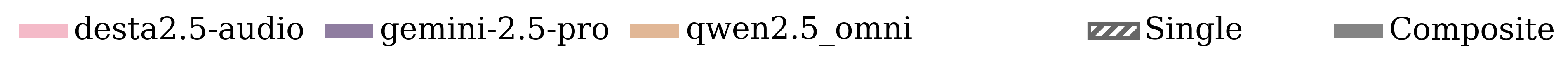}
    
    \setlength{\tabcolsep}{1pt} 
    \begin{tabular}{ccc}
        \includegraphics[height=1.6cm]{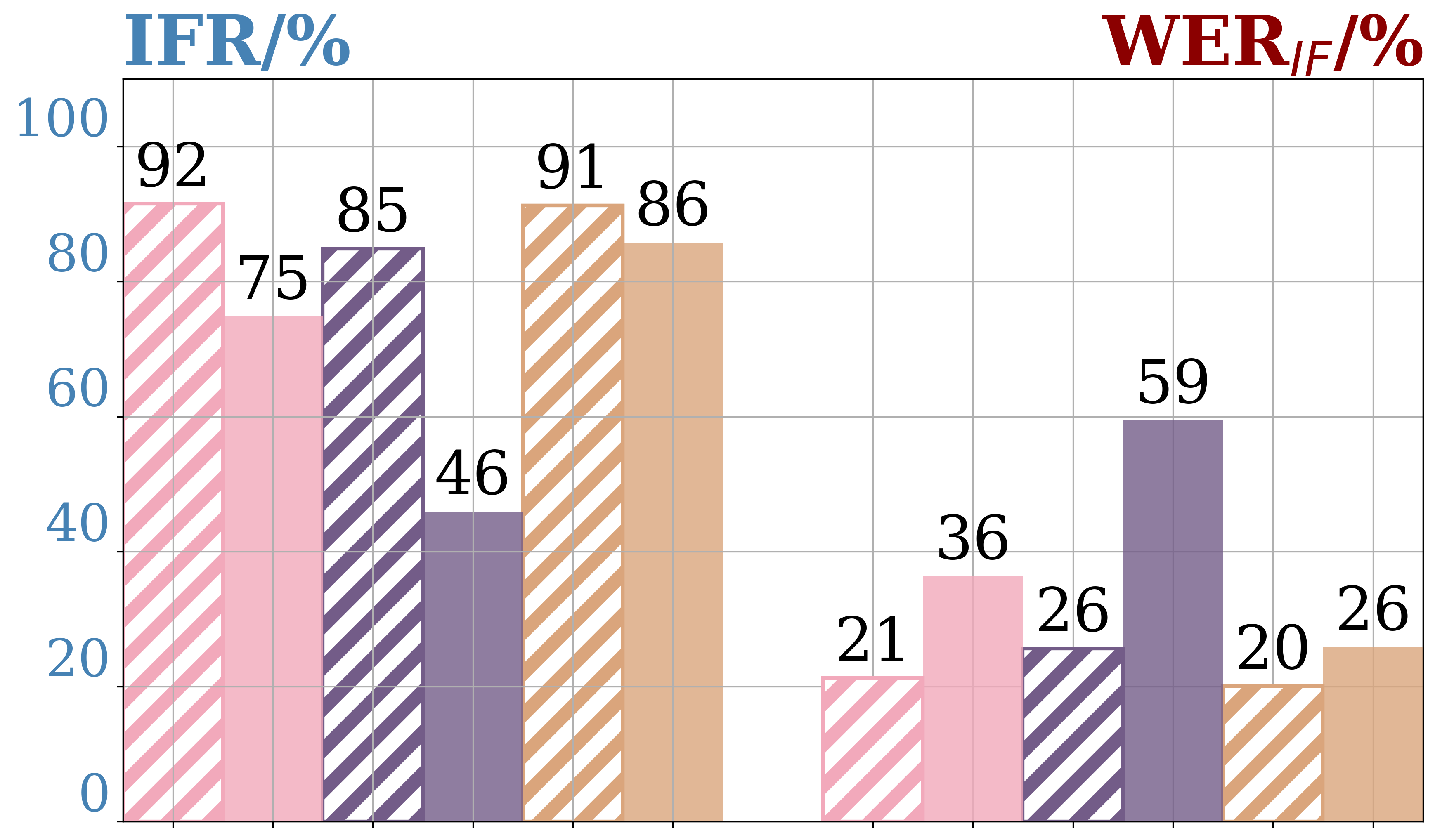} &
        \includegraphics[height=1.6cm]{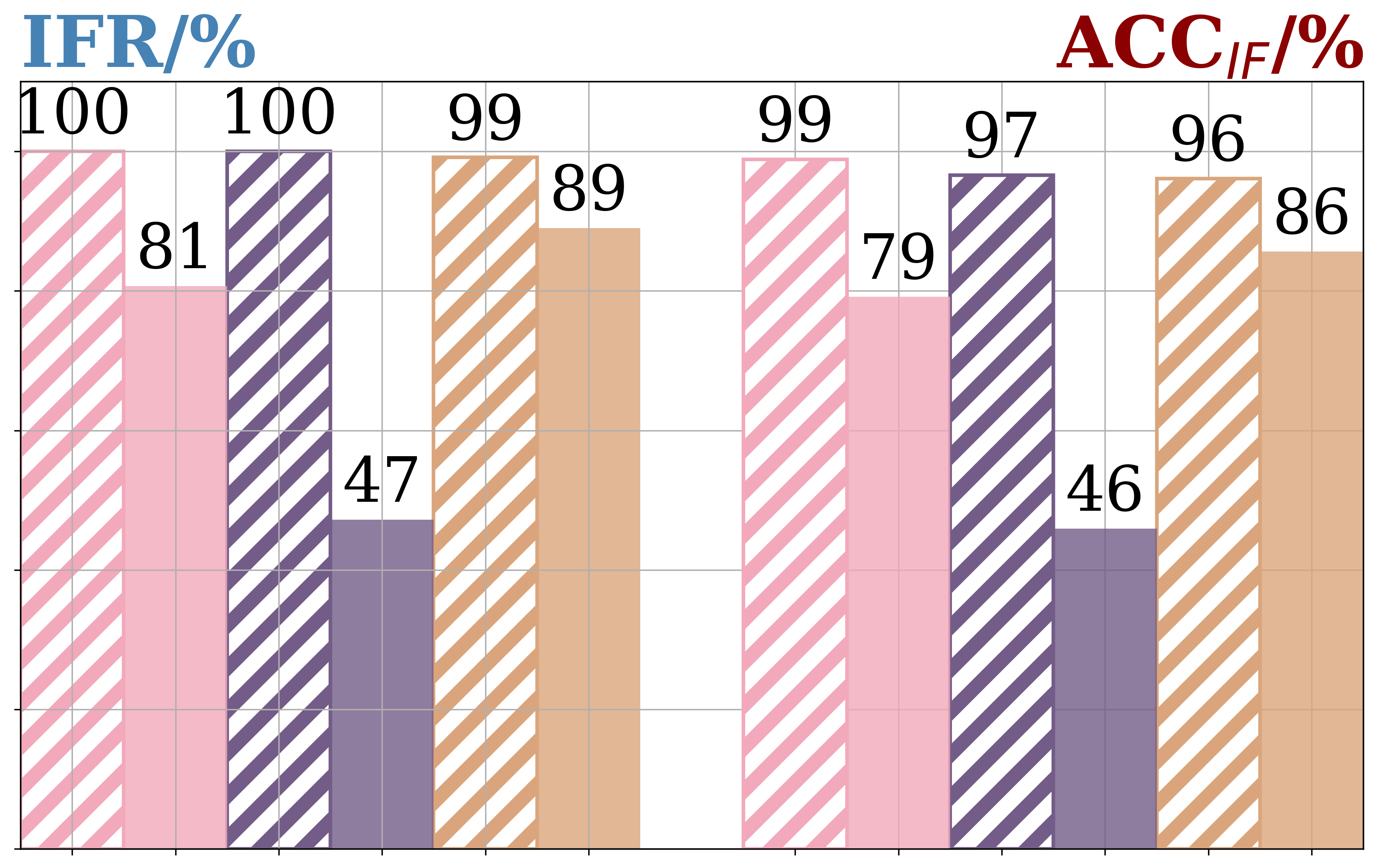} &
        \includegraphics[height=1.6cm]{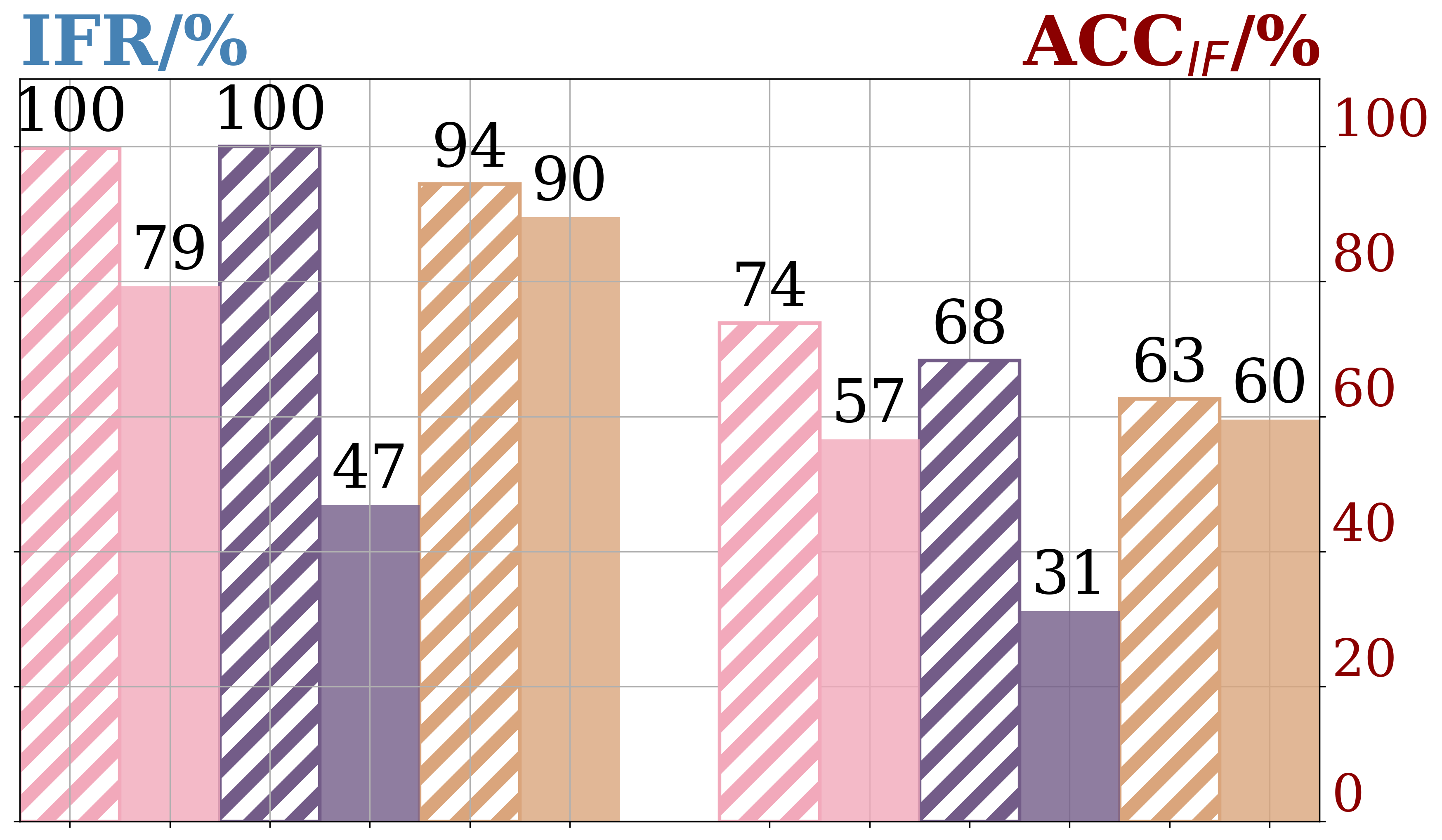} \\
        {\scriptsize ASR} & {\scriptsize GR} & {\scriptsize SER} \\
    \end{tabular}
    \vspace{-10pt}
    \caption{Three models performance on single and composite settings.}
    \label{fig:multi_model_comparison}
    \vspace{-20pt}
\end{figure}

\noindent\textit{\textbf{Zoomed-in Analysis of Task Performance}}\quad We report several the best-performing LALMs for a more detailed task-wise breakdown. As shown in Figure \ref{fig:multi_model_comparison}, when focusing on three atomic tasks—ASR, AAC, and SER—in the \D, \F dimensions, no model consistently excels across all three tasks and both dimensions. For an instance, Qwen2-Audio perform the best on ASR task in \D dimension, while perform the worst in \F dimension. This reveals clear sensitivity to instruction description and formatting requirements: performance varies markedly depending on task type, instruction phrasing and response format. 
For \N dimension, we measure the single task performance with an answer-only constrain requirement. In composite tasks, we apply the same requirement plus json or separation formatting check. We observe that all of the models have degradations on performance of atomic tasks in composite settings. The results indict that composite task basically caused from the decline of instruction following ability.

\begin{figure}[t]
    \centering
    \includegraphics[width=0.95\columnwidth]{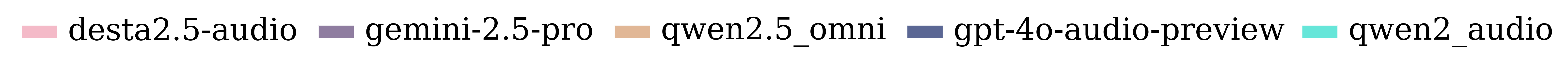}
    \setlength{\tabcolsep}{1pt} 
    \begin{tabular}{ccc}
        \includegraphics[height=1.5cm]{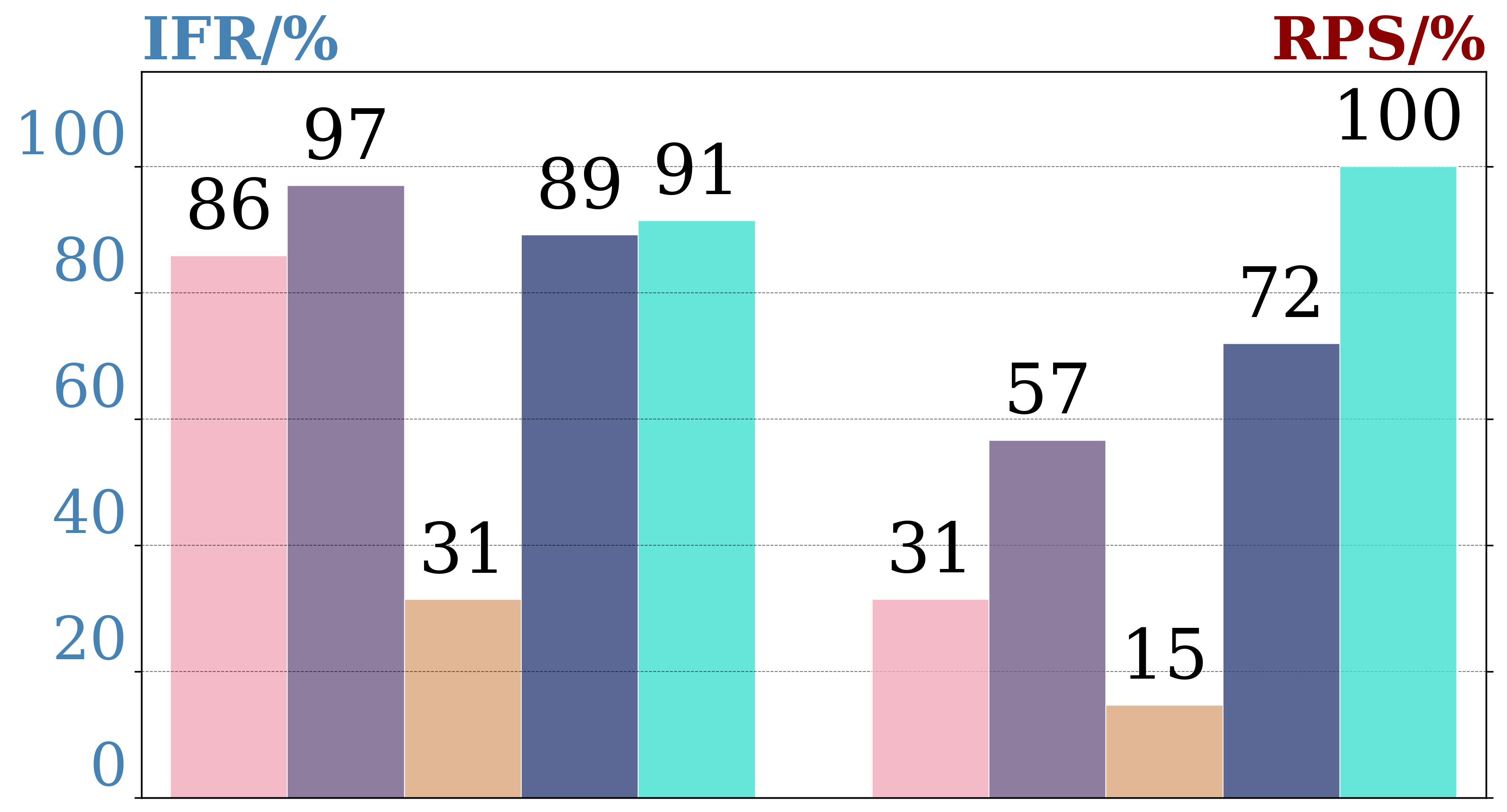} &
        \includegraphics[height=1.5cm]{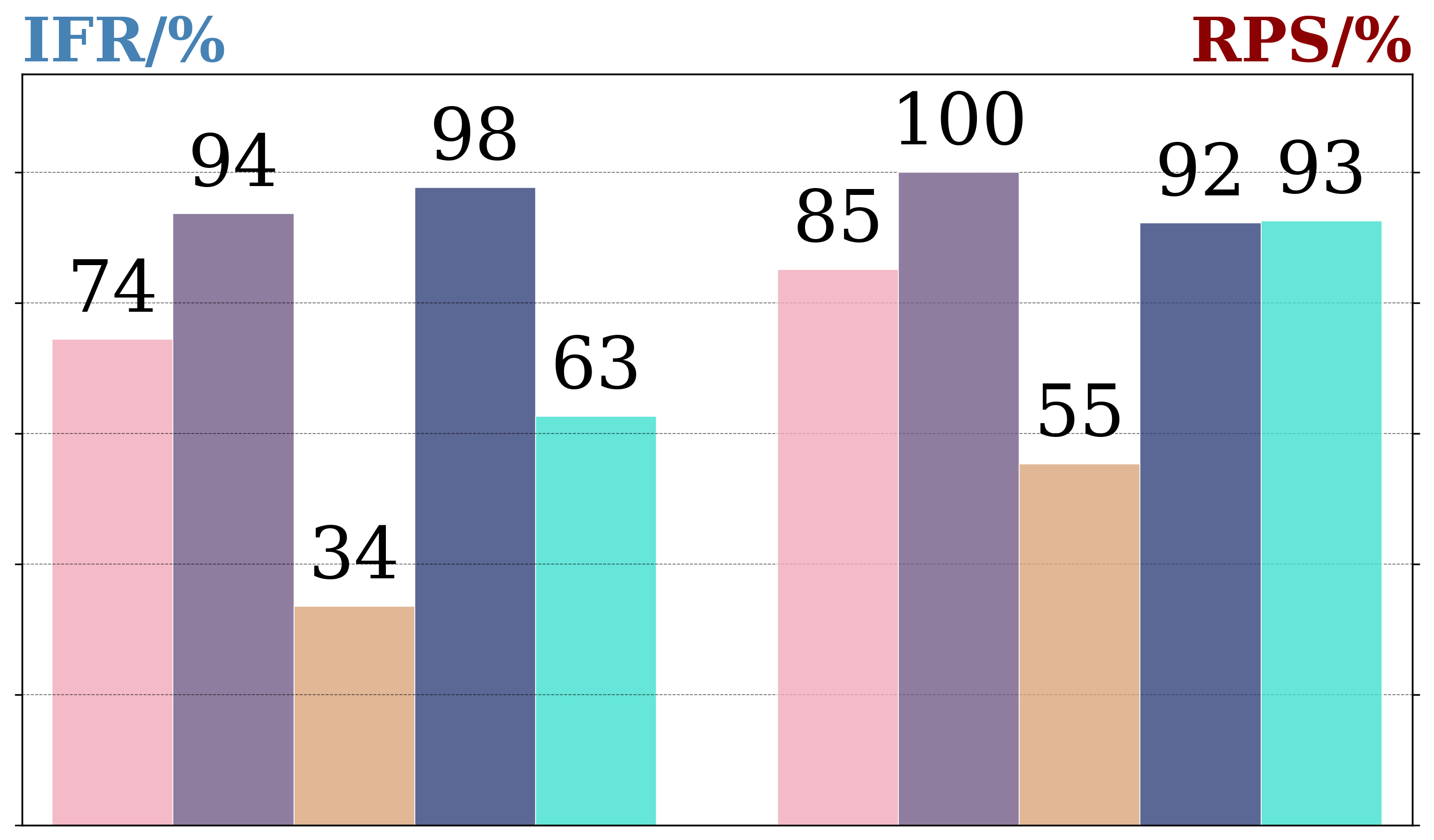} &
        \includegraphics[height=1.5cm]{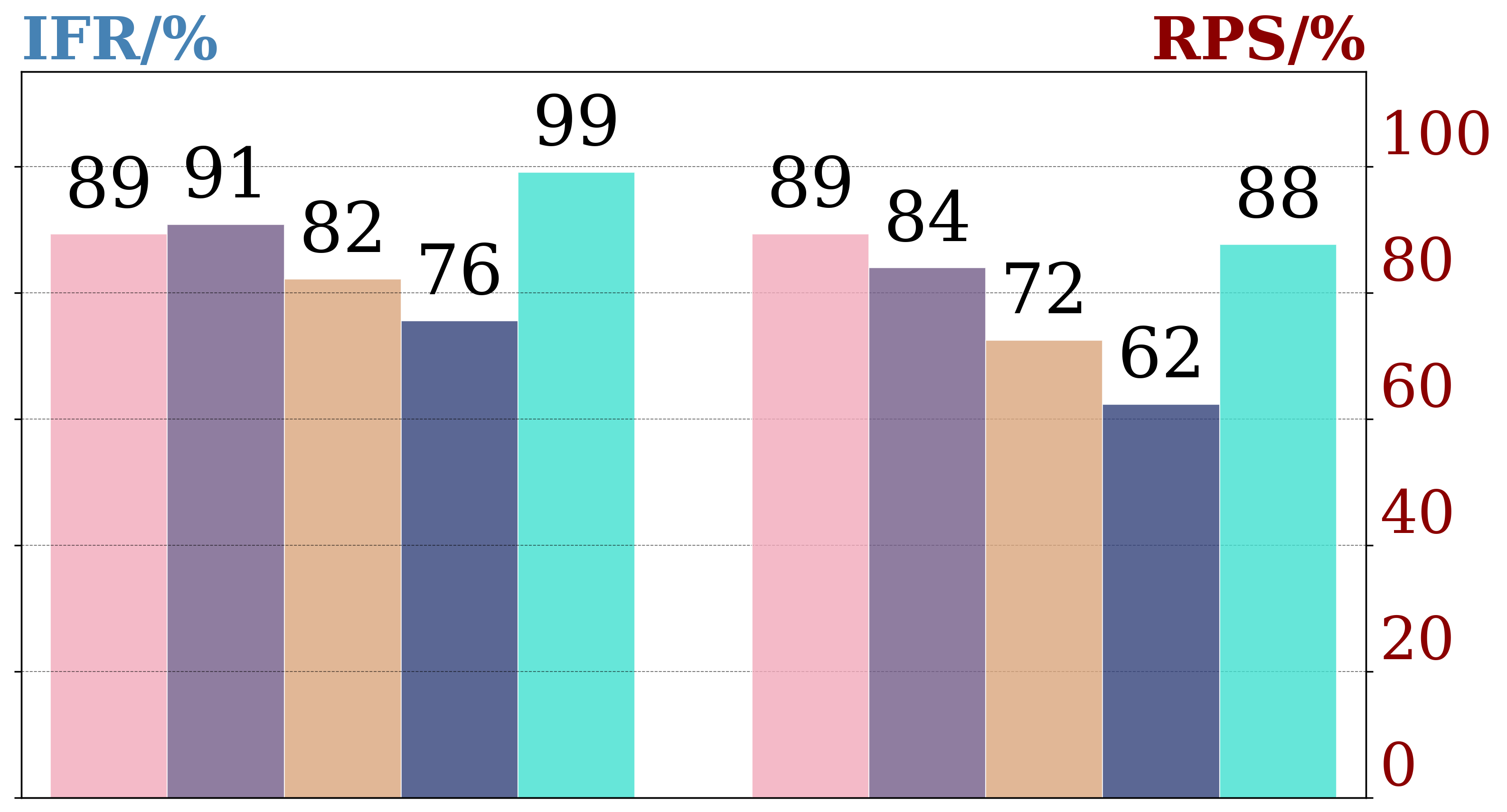} \\
        {\scriptsize ASR (\D)} & {\scriptsize AAC (\D)} & {\scriptsize SER (\D)} \\
    \end{tabular}
    
    
    \setlength{\tabcolsep}{1pt}
    \begin{tabular}{ccc}
        \includegraphics[height=1.5cm]{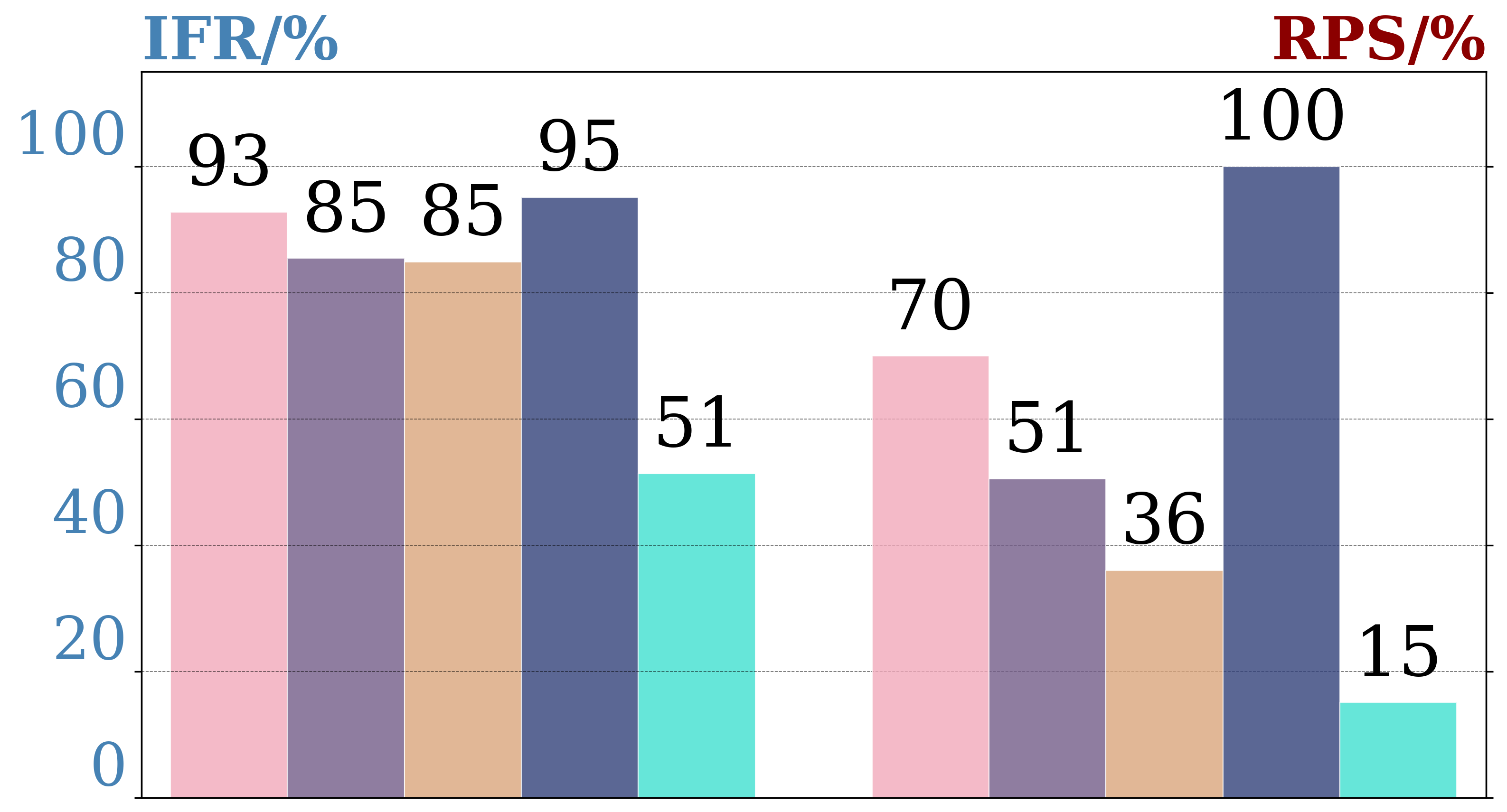} &
        \includegraphics[height=1.5cm]{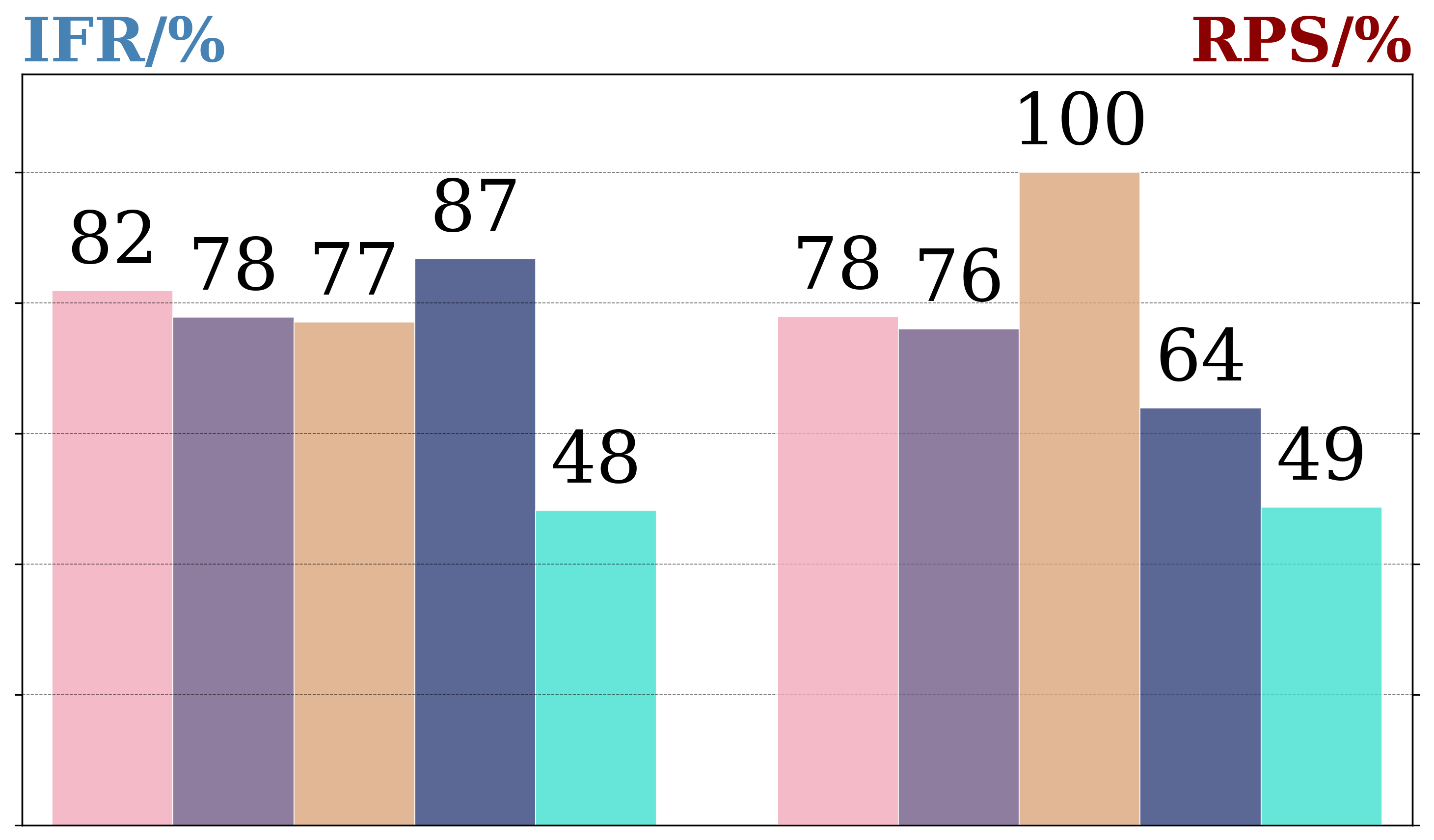} &
        \includegraphics[height=1.5cm]{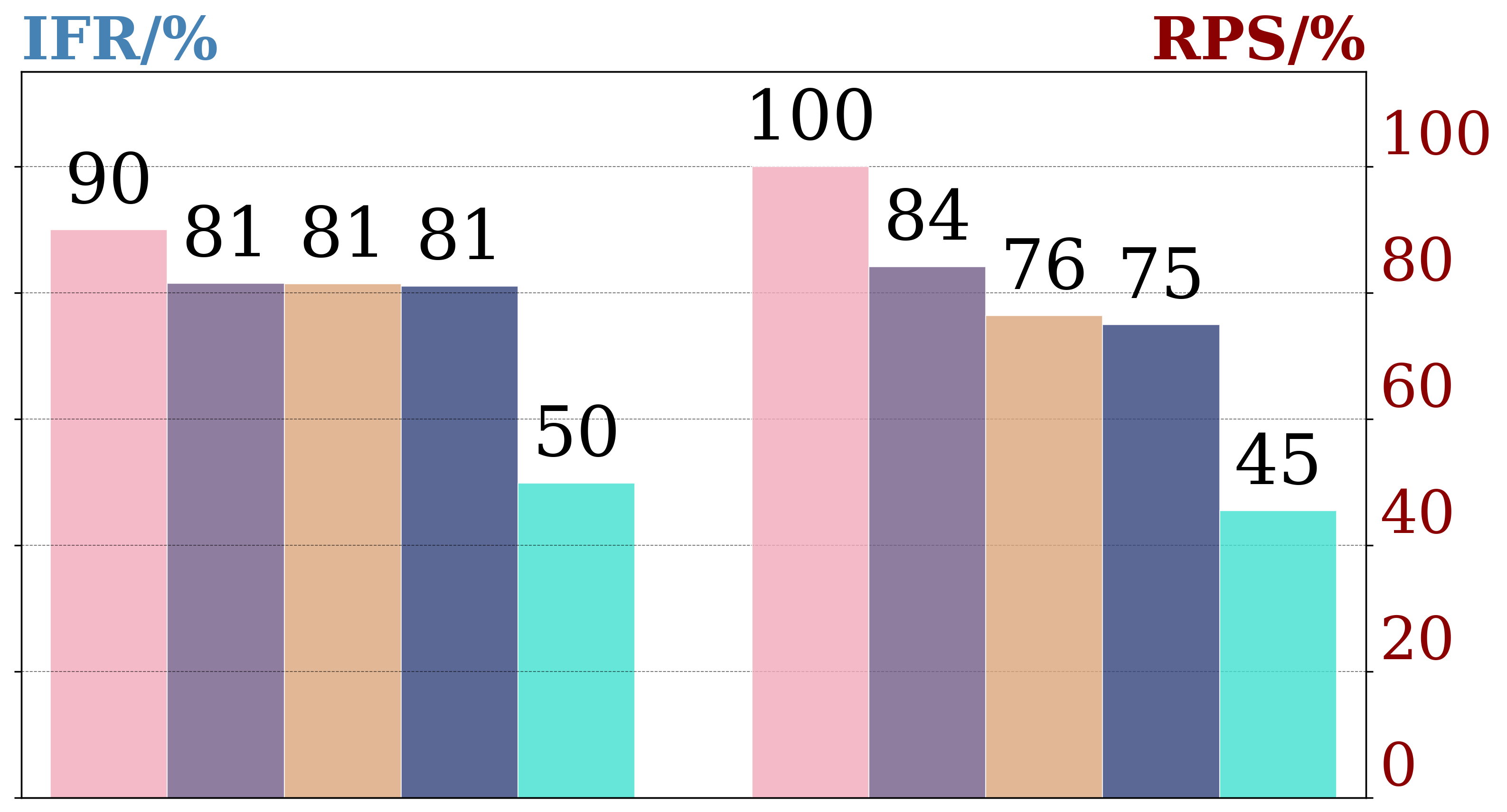} \\
        {\scriptsize ASR (\F)} & {\scriptsize AAC (\F)} & {\scriptsize SER (\F)} \\
    \end{tabular}
    \vspace{-10pt}
    \caption{Five models performance on ASR, AAC, and SER tasks in \D, \F dimensions.}
    \label{fig:three_tasks}
    \vspace{-10pt}
\end{figure}

\vspace{-5pt}

\subsection{Discussion of Supervised Fine-tuning Effectiveness}
We perform mitigation experiments on Qwen2-Audio. Training data are constructed from the corresponding training subsets of the test sets. For SFT in \D and \F dimensions, we generate diverse instruction-variant samples, varying textual descriptions and formatting. Due to limited data, for the \N dimension we conduct SFT separately using samples from IEMOCAP sessions 1–4. As a result, the average instruction-following rates of tested atomic tasks can increase by approximately 9$\%$ under \D, 56$\%$ under \F, and even 
2$\times$ under \N dimension. However, models might suffer catastrophic forgetting cases: they lose previously mastered capabilities when fine-tuned on new instruction variants, only reproducing a few responses similar to those seen during training( \D$\&$\F) or refusing to answer( \N). This indicates simple SFT is insufficient. Strategies such as those employed in DeSTA2.5-Audio~\cite{lu2025desta25audiogeneralpurposelargeaudio} or scaling to much larger, diverse data sets may be considered for instruction sensitivity improvement.

\vspace{-10pt}
\section{Conclusion}
\vspace{-10pt}

In conclusion, we introduce ISA-Bench, a dynamic and comprehensive benchmark for evaluating instruction sensitivity in LALMs. Through diverse instruction variations, carefully designed evaluation strategies, and extensive experiments, this benchmark reveals the significant challenges posed by instruction sensitivity. In particular, our mitigation experiments show that simple supervised fine-tuning (SFT) is insufficient. We hope ISA-Bench will provide valuable insights toward developing robust, human-interaction-friendly LALMs.

\bibliographystyle{IEEEbib}
\bibliography{refs}

\end{document}